\documentclass[sigconf]{aamas} 

\usepackage{balance} 
\usepackage{soul}
\usepackage{caption}
\usepackage{booktabs}
\usepackage[normalem]{ulem}
\usepackage{MnSymbol}
\usepackage{stmaryrd}
\usepackage{graphicx}
\usepackage{amsmath}

\newcommand{\joy}[3]{\ensuremath{\operatorname{joy}_{#1}(#2,#3)}}
\newcommand{\joyl}[2]{\ensuremath{\operatorname{ijoy}_{#1}(#2)}}

\numberwithin{definition}{section}
\numberwithin{proposition}{section}
\numberwithin{theorem}{section}
\numberwithin{assumption}{section}
\numberwithin{CON}{section}

\setcopyright{ifaamas}
\acmConference[AAMAS '23]{Proc.\@ of the 22nd International Conference
on Autonomous Agents and Multiagent Systems (AAMAS 2023)}{May 29 -- June 2, 2023}
{London, United Kingdom}{A.~Ricci, W.~Yeoh, N.~Agmon, B.~An (eds.)}
\copyrightyear{2023}
\acmYear{2023}
\acmDOI{}
\acmPrice{}
\acmISBN{}

\acmSubmissionID{262}

\title[AAMAS-2023 Formatting Instructions]{What Do You Care About: Inferring Values from Emotions}
\subtitle{Extended Abstract}

\author{Jieting Luo}
\affiliation{
  \institution{University of Bern}
  \city{Bern}
  \country{Switzerland}}
\email{jieting.luo@unibe.ch}

\author{Mehdi Dastani}
\affiliation{
  \institution{Utrecht University}
  \city{Utrecht}
  \country{Netherlands}}
\email{M.M.Dastani@uu.nl}

\author{Thomas Studer}
\affiliation{
  \institution{University of Bern}
  \city{Bern}
  \country{Switzerland}}
\email{thomas.studer@unibe.ch}

\author{Beishui Liao}
\affiliation{
  \institution{Zhejiang University}
  \city{Hangzhou}
  \country{China}}
\email{baiseliao@zju.edu.cn}

\begin{abstract}
Observers can glean information from others’ emotional expressions through the act of drawing inferences from another individual’s emotional expressions. It is important for socially aware artificial systems to be capable of doing that as it can facilitate social interaction among agents, and is particularly important in human-robot interaction for supporting a more personalized treatment of users. In this short paper, we propose a methodology for developing a formal model that allows agents to infer another agent's values from her emotion expressions.
\end{abstract}


\newcommand{\BibTeX}{\rm B\kern-.05em{\sc i\kern-.025em b}\kern-.08em\TeX}

\begin{document}


\pagestyle{fancy}
\fancyhead{}


\maketitle 


\section{Introduction}
As it has been recognized that emotions play an important and functional role in driving individuals' behavior, some efforts have been made in artificial intelligence to provide formal models of emotions \cite{dastani2012logic,dastani2017other,adam2009logical,steunebrink2012formal,steunebrink2009formal}. Another important dimension of emotions is the social aspect, which concerns emotion expression of one agent influences the behavior of another agent with whom it interacts. In particular, observers can glean information from others’ emotional expressions through a process that is called inferential processes by van Kleef in \cite{van2009emotions}. Such information includes the inner state of the agent who expresses emotions, her social intentions, and her orientation toward other agents. For instance, if an agent believes that another agent is joyful for buying ice cream, the observer agent will infer that the expresser agent likes ice cream. It is important to implement the mechanism of inferential processes in socially aware artificial systems as it can facilitate social interaction among agents, and is particularly important in human-robot interaction for supporting a more personalized treatment of users.

However, inferring an agent's values from her emotion expressions is challenging \cite{ong2019computational}. Current emotion-recognition approaches based on machine learning are able to distinguish different types of emotions from facial expressions, but are not capable of inferring the agent's inner state that causes the emotion \cite{dzedzickis2020human}\cite{shu2018review}. One intuitive approach to this problem is to allow the observer agent to connect an observed emotion with her background knowledge about the event that has triggered the emotion. For instance, if an agent believes that another agent is joyful after buying ice cream, and the observer agent has the belief that buying ice cream will result in having ice cream, she can infer that the observed agent desires to have ice cream. In this paper, we propose a methodology for developing a formal model that allow agents to infer another agent's values from her emotions expressions. It is essential in understanding and modeling the inferential process of emotions in social settings and paves the way for the design and development of software agents that are able to establish affective interaction with human users.

\section{Logical Framework}\label{sec:model}
As we aim at modeling interaction scenarios, we consider a multi-agent setting consisting of agents, actions and states. A transition relation represents the dynamics of the system and it is labeled with an agent's action. An epistemic accessibility relation for each agent $i$ is used to represent agents' belief. According to appraisal theory in psychology, emotions result from people's evaluation of their perceived events. Different agents may have different emotional reactions to the same event, depending on their standards they use for evaluation. For example, kids gets joyful for having dessert while diabetes patients gets distressed for having dessert. Thus, we can use personal values as agents' evaluation standard, which are similar to goals and desires. In order to simplify our methodology, we assume that a value $v$ is a literal in a finite set $L$ and agent $i$'s procession of value $v$, expressed as $\operatorname{Val}_i(v))$, is a state property. Having this framework allows agents to reason about actions, beliefs of the environment and another agent's procession of values.

\section{Modeling Emotions}\label{sec:emotions}
In order to model the emotion theory that agents share for their affective interaction, we adopt the OCC psychological model of emotions where emotions are structured based on \emph{focus of attention} \cite{ortony1990cognitive}. Values are used as an agent's internal evaluation standard as in \cite{luo2022modeling}. With this psychological model of emotions, we can define an agent $i$'s emotion with respect to an action $a$ and a value $v$ as a state property. For example, 
\[\joy{i}{a}{v} \overset{def}{=} B_i (v \land \langle -a \rangle \lnot v) \land \operatorname{Val}_i(v), \]
which means that agent $i$ is joyful with respect to action $a$ and value $v$ if and only if agent $i$ believes that it is the case that $v$ holds and $v$ did not hold before action $a$ was performed and agent $i$ has value $v$. An emotion contains important information about expressers' belief of the environment and their values based on which they evaluate states and actions. In reality, however, it is possible that the observer agent observes an emotion while the reason behind is not clear. For example, an agent observes that her user is distressed for dropping a cup of coffee, but the agent is not sure what the user cares about in this incident. This is called an incomplete emotion in this paper. For example, 
\[\joyl{i}{a} \overset{def}{=} \underset{v \in L}{\bigvee}(B_i (v \land \langle -a \rangle \lnot v) \land \operatorname{Val}_i(v)), \]
which means that agent $i$ is joyful with respect to action $a$ if only if there exists a literal $v$ in $L$ such that agent $i$ believes it is true and it was false before action $a$ was performed and $v$ is agent $i$'s value. 

\section{Inferential Process}\label{inference}
If an agent observes and believes the emotion expression with complete information from another agent, the observer agent can infer the expresser's belief of the environment and her value based on which she evaluates states and actions. For instance, if an agent believes that another agent is distressed about getting rice, the former agent can infer that the latter agent does not like eating rice; if an agent believes that another agent is hopeful for getting her paper accepted, the former agent can infer that paper acceptance is important for the latter agent. Such a process is called \emph{inferential processes} in \cite{van2016interpersonal}. What if the observer agent observes an incomplete emotion? The action that have triggered the emotion is given, but the consequence of the action that the emotion is directed at is unknown. In such a case, the observer can infer this internal information through connecting the observed incomplete emotion with her background knowledge about the consequence of action. For example, a robot observes that her user gets distressed for dropping a cup of coffee, and from this incident the robot can infer that her user cares about the cup or the coffee inside the cup. An incomplete emotion results in disjunctions, which is a form of ambiguity, but \emph{new} information that is inferred over time can be used to exclude some disjuncts. If the robot continues to observe her user's emotions, one day she might find out that her user is not distressed for dropping a cup without coffee inside. From this the robot can infer that the user does not care about the cup, and he can conclude based on previous information that the user cares about the coffee. As we can see, the observer agent infers and collects useful information from her observation history, and thus she needs an approach to handle all this information for disambiguating another agent's values. 

Formal argumentation is a nonmonotonic formalism for representing and reasoning about conflicts based on the construction and the evaluation of interacting arguments \cite{dung1995acceptability}. If two arguments have opposite conclusions, then they attack each other. A maximal consistent set of arguments can be found using Dung's well-known argumentation semantics. In particular, structured argumentation frameworks such as ASPIC$^+$ and ABA allows us to build an argumentation theory on the meta level over classical logic \cite{modgil2013general}\cite{toni2014tutorial}. We thus propose to use formal argumentation to resolve conflicts between arguments as a way for the observer agent to disambiguate another agent's values. An observer agent constructs an ordinary argument, supporting a value $v$ belongs to an expresser agent, when the observer agent believes an incomplete emotion with respect to an action from the expresser agent and that the action usually brings about a consequence $v$. An observer agent can also construct a blocking argument, opposing a value $v$ belongs to an expresser agent, when the observer agent believes that the expresser agent does not express a specific incomplete emotion with respect to an action but believes that the action usually brings about a consequence $v$. The observer agent can then specify the attack relation based on what an argument supports and opposes. Following the interpretation of the arguments, two ordinary arguments attacks each other if and only if they support different values for an incomplete emotion that is observed in the same state, because only one emotion is observed in a state and we do not allow an agent to express an emotion for two different values given that a value is assumed to be a literal. One blocking argument attacks one ordinary argument if and only if the value that the blocking argument opposes is supported by the ordinary argument. We now can construct a Dung-style abstract argumentation framework AFV with respect to an observation history and an agent's values. Because an AFV is defined with respect to an observation history, once the observer agent performs one more observation, she can construct more arguments regarding the expresser agent's values and more attacks and add them to the current AFV. Under the grounded semantic, the observer agent can calculate a set of arguments that are critically acceptable together, which indicate a set of values that are believed to belong to the expresser agent and a set of values that are believed not to belong to the expresser agent. But it should be emphasized that the result only holds with respect to the observation history, because it might change as the observer agent constructs more arguments due to her further observations and updates the argumentation framework accordingly.

\section{Conclusions}
This paper proposes a methodology for modeling the mechanism of influential processes that allows agents to infer another agent's values through observing her emotional expressions. It can be used for building agents' Theory of Mind, which refers to the capacity to understand other individuals by ascribing mental states to them. Moreover, it can be extended for the inference of social and ethical values among a society, which has close connection with social norms, and thus future work can be done on the identification of social norms from emotion expressions based on our model. Another direction can be to enrich and advance our model for inferring another agent’s value \emph{preferences} apart from her procession of values from her emotion expressions.

\begin{acks}
Jieting Luo and Thomas Studer are supported by the Swiss National Science Foundation grant 200020$\_$184625.
\end{acks}



\bibliographystyle{ACM-Reference-Format} 
\bibliography{sigproc}


\begin{thebibliography}{15}


\ifx \showCODEN    \undefined \def \showCODEN     #1{\unskip}     \fi
\ifx \showDOI      \undefined \def \showDOI       #1{#1}\fi
\ifx \showISBNx    \undefined \def \showISBNx     #1{\unskip}     \fi
\ifx \showISBNxiii \undefined \def \showISBNxiii  #1{\unskip}     \fi
\ifx \showISSN     \undefined \def \showISSN      #1{\unskip}     \fi
\ifx \showLCCN     \undefined \def \showLCCN      #1{\unskip}     \fi
\ifx \shownote     \undefined \def \shownote      #1{#1}          \fi
\ifx \showarticletitle \undefined \def \showarticletitle #1{#1}   \fi
\ifx \showURL      \undefined \def \showURL       {\relax}        \fi
\providecommand\bibfield[2]{#2}
\providecommand\bibinfo[2]{#2}
\providecommand\natexlab[1]{#1}
\providecommand\showeprint[2][]{arXiv:#2}

\bibitem[\protect\citeauthoryear{Adam, Herzig, and Longin}{Adam
  et~al\mbox{.}}{2009}]%
        {adam2009logical}
\bibfield{author}{\bibinfo{person}{Carole Adam}, \bibinfo{person}{Andreas
  Herzig}, {and} \bibinfo{person}{Dominique Longin}.}
  \bibinfo{year}{2009}\natexlab{}.
\newblock \showarticletitle{A logical formalization of the OCC theory of
  emotions}.
\newblock \bibinfo{journal}{\emph{Synthese}} \bibinfo{volume}{168},
  \bibinfo{number}{2} (\bibinfo{year}{2009}), \bibinfo{pages}{201--248}.
\newblock


\bibitem[\protect\citeauthoryear{Dastani and Lorini}{Dastani and
  Lorini}{2012}]%
        {dastani2012logic}
\bibfield{author}{\bibinfo{person}{Mehdi Dastani} {and}
  \bibinfo{person}{Emiliano Lorini}.} \bibinfo{year}{2012}\natexlab{}.
\newblock \showarticletitle{A logic of emotions: from appraisal to coping.}. In
  \bibinfo{booktitle}{\emph{AAMAS}}. \bibinfo{pages}{1133--1140}.
\newblock


\bibitem[\protect\citeauthoryear{Dastani, Lorini, Meyer, and Pankov}{Dastani
  et~al\mbox{.}}{2017}]%
        {dastani2017other}
\bibfield{author}{\bibinfo{person}{Mehdi Dastani}, \bibinfo{person}{Emiliano
  Lorini}, \bibinfo{person}{John-Jules Meyer}, {and} \bibinfo{person}{Alexander
  Pankov}.} \bibinfo{year}{2017}\natexlab{}.
\newblock \showarticletitle{Other-condemning anger= blaming accountable agents
  for unattainable desires}. In \bibinfo{booktitle}{\emph{International
  Conference on Principles and Practice of Multi-Agent Systems}}. Springer,
  \bibinfo{pages}{15--33}.
\newblock


\bibitem[\protect\citeauthoryear{Dung}{Dung}{1995}]%
        {dung1995acceptability}
\bibfield{author}{\bibinfo{person}{Phan~Minh Dung}.}
  \bibinfo{year}{1995}\natexlab{}.
\newblock \showarticletitle{On the acceptability of arguments and its
  fundamental role in nonmonotonic reasoning, logic programming and n-person
  games}.
\newblock \bibinfo{journal}{\emph{Artificial intelligence}}
  \bibinfo{volume}{77}, \bibinfo{number}{2} (\bibinfo{year}{1995}),
  \bibinfo{pages}{321--357}.
\newblock


\bibitem[\protect\citeauthoryear{Dzedzickis, Kaklauskas, and
  Bucinskas}{Dzedzickis et~al\mbox{.}}{2020}]%
        {dzedzickis2020human}
\bibfield{author}{\bibinfo{person}{Andrius Dzedzickis},
  \bibinfo{person}{Art{\=u}ras Kaklauskas}, {and} \bibinfo{person}{Vytautas
  Bucinskas}.} \bibinfo{year}{2020}\natexlab{}.
\newblock \showarticletitle{Human emotion recognition: Review of sensors and
  methods}.
\newblock \bibinfo{journal}{\emph{Sensors}} \bibinfo{volume}{20},
  \bibinfo{number}{3} (\bibinfo{year}{2020}), \bibinfo{pages}{592}.
\newblock


\bibitem[\protect\citeauthoryear{Luo and Dastani}{Luo and Dastani}{2022}]%
        {luo2022modeling}
\bibfield{author}{\bibinfo{person}{Jieting Luo} {and} \bibinfo{person}{Mehdi
  Dastani}.} \bibinfo{year}{2022}\natexlab{}.
\newblock \showarticletitle{Modeling Affective Reaction in Multi-agent
  Systems}. In \bibinfo{booktitle}{\emph{Proceedings of the 21st International
  Conference on Autonomous Agents and Multiagent Systems}}.
  \bibinfo{pages}{1681--1683}.
\newblock


\bibitem[\protect\citeauthoryear{Modgil and Prakken}{Modgil and
  Prakken}{2013}]%
        {modgil2013general}
\bibfield{author}{\bibinfo{person}{Sanjay Modgil} {and} \bibinfo{person}{Henry
  Prakken}.} \bibinfo{year}{2013}\natexlab{}.
\newblock \showarticletitle{A general account of argumentation with
  preferences}.
\newblock \bibinfo{journal}{\emph{Artificial Intelligence}}
  \bibinfo{volume}{195} (\bibinfo{year}{2013}), \bibinfo{pages}{361--397}.
\newblock


\bibitem[\protect\citeauthoryear{Ong, Zaki, and Goodman}{Ong
  et~al\mbox{.}}{2019}]%
        {ong2019computational}
\bibfield{author}{\bibinfo{person}{Desmond~C Ong}, \bibinfo{person}{Jamil
  Zaki}, {and} \bibinfo{person}{Noah~D Goodman}.}
  \bibinfo{year}{2019}\natexlab{}.
\newblock \showarticletitle{Computational models of emotion inference in theory
  of mind: A review and roadmap}.
\newblock \bibinfo{journal}{\emph{Topics in cognitive science}}
  \bibinfo{volume}{11}, \bibinfo{number}{2} (\bibinfo{year}{2019}),
  \bibinfo{pages}{338--357}.
\newblock


\bibitem[\protect\citeauthoryear{Ortony, Clore, and Collins}{Ortony
  et~al\mbox{.}}{1990}]%
        {ortony1990cognitive}
\bibfield{author}{\bibinfo{person}{Andrew Ortony}, \bibinfo{person}{Gerald~L
  Clore}, {and} \bibinfo{person}{Allan Collins}.}
  \bibinfo{year}{1990}\natexlab{}.
\newblock \bibinfo{booktitle}{\emph{The cognitive structure of emotions}}.
\newblock \bibinfo{publisher}{Cambridge university press}.
\newblock


\bibitem[\protect\citeauthoryear{Shu, Xie, Yang, Li, Li, Liao, Xu, and
  Yang}{Shu et~al\mbox{.}}{2018}]%
        {shu2018review}
\bibfield{author}{\bibinfo{person}{Lin Shu}, \bibinfo{person}{Jinyan Xie},
  \bibinfo{person}{Mingyue Yang}, \bibinfo{person}{Ziyi Li},
  \bibinfo{person}{Zhenqi Li}, \bibinfo{person}{Dan Liao},
  \bibinfo{person}{Xiangmin Xu}, {and} \bibinfo{person}{Xinyi Yang}.}
  \bibinfo{year}{2018}\natexlab{}.
\newblock \showarticletitle{A review of emotion recognition using physiological
  signals}.
\newblock \bibinfo{journal}{\emph{Sensors}} \bibinfo{volume}{18},
  \bibinfo{number}{7} (\bibinfo{year}{2018}), \bibinfo{pages}{2074}.
\newblock


\bibitem[\protect\citeauthoryear{Steunebrink, Dastani, and Meyer}{Steunebrink
  et~al\mbox{.}}{2009}]%
        {steunebrink2009formal}
\bibfield{author}{\bibinfo{person}{Bas~R Steunebrink}, \bibinfo{person}{Mehdi
  Dastani}, {and} \bibinfo{person}{John-Jules~Ch Meyer}.}
  \bibinfo{year}{2009}\natexlab{}.
\newblock \showarticletitle{A formal model of emotion-based action tendency for
  intelligent agents}. In \bibinfo{booktitle}{\emph{Progress in Artificial
  Intelligence: 14th Portuguese Conference on Artificial Intelligence, EPIA
  2009, Aveiro, Portugal, October 12-15, 2009. Proceedings 14}}. Springer,
  \bibinfo{pages}{174--186}.
\newblock


\bibitem[\protect\citeauthoryear{Steunebrink, Dastani, and Meyer}{Steunebrink
  et~al\mbox{.}}{2012}]%
        {steunebrink2012formal}
\bibfield{author}{\bibinfo{person}{Bas~R Steunebrink}, \bibinfo{person}{Mehdi
  Dastani}, {and} \bibinfo{person}{John-Jules~Ch Meyer}.}
  \bibinfo{year}{2012}\natexlab{}.
\newblock \showarticletitle{A formal model of emotion triggers: an approach for
  BDI agents}.
\newblock \bibinfo{journal}{\emph{Synthese}} \bibinfo{volume}{185},
  \bibinfo{number}{1} (\bibinfo{year}{2012}), \bibinfo{pages}{83--129}.
\newblock


\bibitem[\protect\citeauthoryear{Toni}{Toni}{2014}]%
        {toni2014tutorial}
\bibfield{author}{\bibinfo{person}{Francesca Toni}.}
  \bibinfo{year}{2014}\natexlab{}.
\newblock \showarticletitle{A tutorial on assumption-based argumentation}.
\newblock \bibinfo{journal}{\emph{Argument \& Computation}}
  \bibinfo{volume}{5}, \bibinfo{number}{1} (\bibinfo{year}{2014}),
  \bibinfo{pages}{89--117}.
\newblock


\bibitem[\protect\citeauthoryear{Van~Kleef}{Van~Kleef}{2009}]%
        {van2009emotions}
\bibfield{author}{\bibinfo{person}{Gerben~A Van~Kleef}.}
  \bibinfo{year}{2009}\natexlab{}.
\newblock \showarticletitle{How emotions regulate social life: The emotions as
  social information (EASI) model}.
\newblock \bibinfo{journal}{\emph{Current directions in psychological science}}
  \bibinfo{volume}{18}, \bibinfo{number}{3} (\bibinfo{year}{2009}),
  \bibinfo{pages}{184--188}.
\newblock


\bibitem[\protect\citeauthoryear{Van~Kleef}{Van~Kleef}{2016}]%
        {van2016interpersonal}
\bibfield{author}{\bibinfo{person}{Gerben~A Van~Kleef}.}
  \bibinfo{year}{2016}\natexlab{}.
\newblock \bibinfo{booktitle}{\emph{The interpersonal dynamics of emotion}}.
\newblock \bibinfo{publisher}{Cambridge University Press}.
\newblock


\end{thebibliography}


\end{document}